\documentclass[journal]{IEEEtran}
\ifCLASSINFOpdf
\else
\fi
\usepackage{graphics} 
\usepackage{epsfig} 
\usepackage{balance}
\usepackage{yhmath} 
\usepackage{color}

\usepackage{times} 
\usepackage{amsmath} 
\usepackage{amssymb}  
\usepackage{algorithm}
\usepackage{algorithmic}

\newtheorem{lemma}{Lemma}

\newtheorem{remark}{Remark}

\begin{document}
%
\title{A Connectivity-Aware Approximation Algorithm for Relay Node Placement in Wireless Sensor Networks}
%
%
%

\author{Chaofan Ma,
        Wei Liang$^*$,
        Meng Zheng, and Hamid Sharif
\thanks{This work was supported by the Natural Science Foundation of China (61233007 and 61172145) and Cross-disciplinary Collaborative Teams Program for Science, Technology and Innovation of Chinese Academy of Sciences (Network and System Technologies for Safety Monitoring and Information Interacting in Smart Grid). Part of this paper has been reported in \cite{Ma15}.}
\thanks{Chaofan Ma, Wei Liang, and Meng Zheng are with the Key Laboratory of Networked Control Systems, Chinese Academy of Sciences, Shenyang 110016, China. E-mail: \{machaofan, weiliang, zhengmeng\_6\}@sia.cn.}
\thanks{Chaofan Ma is also with the University of Chinese Academy of Sciences, Beijing 100049, China.}
\thanks{Hamid Sharif is with the Department of Computer \& Electronics Engineering, University of Nebraska-Lincoln, Omaha, NE 68182-05729, USA. Email: hsharif@unl.edu.}
\thanks{$^*$ Corresponding author.}}

\maketitle

\begin{abstract}

In two-tiered Wireless Sensor Networks (WSNs) relay node placement is one of the key factors impacting the network energy consumption and the system overhead. In this paper, a novel connectivity-aware approximation algorithm for relay node placement in WSNs is proposed to offer a major step forward in saving system overhead. Specifically, a unique Local Search Approximation Algorithm (LSAA) is introduced to solve the Relay Node Single Cover (RNSC) problem. In this proposed LSAA approach, the sensor nodes are allocated into groups and then a local Set Cover (SC) for each group is achieved by a local search algorithm. The union set of all local SCs constitutes a SC of the RNSC problem. The approximation ratio and the time complexity of the LSAA are analyzed by rigorous proof. Additionally, the LSAA approach has been extended to solve the relay node double cover problem. Then, a Relay Location Selection Algorithm (RLSA) is proposed to utilize the resulting SC from LSAA in combining RLSA with the minimum spanning tree heuristic to build the high-tier connectivity. As the RLSA searches for a nearest location to the sink node for each relay node, the high-tier network built by RLSA becomes denser than that by existing works. As a result, the number of added relay nodes for building the connectivity of the high-tier WSN can be significantly saved. Simulation results clearly demonstrate that the proposed LSAA outperforms the approaches reported in literature and the RLSA-based algorithm can noticeably save relay nodes newly deployed for the high-tier connectivity.

\end{abstract}

\begin{IEEEkeywords}
wireless sensor networks, relay node placement, geometric disc cover, network connectivity, approximation algorithm.
\end{IEEEkeywords}

%
\IEEEpeerreviewmaketitle

\section{Introduction}
%
%
%
%
\IEEEPARstart{W}{ireless} Sensor Networks (WSNs) have been one of the most important research areas since last decade due to its tremendous application
potentials in civilian and defense related applications such as battlefield surveillance, environmental monitoring, biomedical observation, industrial automation and other fields \cite{Akyildiz02}-\cite{Estrin99}. WSNs typically composed of many low-cost and low-power homogenous or heterogeneous sensor nodes, which can perform sensing,
simple computations, and short-range wireless communications. In many WSN applications, the lifetime of network is limited due to the constraints
in energy resources and accessibility of the actual sensor nodes \cite{Yick08}-\cite{He10}. The two-tiered network architecture has been proposed to
extend the network lifetime in WSNs. In this approach a small number of nodes with ample power and suitable wireless communication radius are
placed to serve as the relay nodes. The relay nodes function like the cluster heads to collect sensed information from one-hop-neighbor sensor
nodes and transmit the data to the sink node \cite{Gupta03}-\cite{He12}. This is an energy-efficient approach since it reduces the
energy and resource consumption of each sensor node and relies on the mesh networks of relay nodes to transmit data from sensor nodes. This highlights
the key role of relay nodes and the importance of investigating their placement considering the required energy consumption and other overheads.

The research problems on relay node placement in two-tiered WSNs can be divided into two subgroups, 1) the Relay Node Cover (RNC), which is mathematically equivalent to the Geometric Disc Covering (GDC) problem, and 2) the network
connectivity (NC). Different published works have investigated the problems of GDC and NC independently, however, this paper considers the two subproblems jointly, i.e., the relay nodes for covering the low-tier WSN are optimally deployed to facilitate the connectivity of the high-tier WSN.

This paper first focuses on the GDC problem and proposes a novel Local Search Approximation Algorithm (LSAA) to the Relay Node Single Cover (RNSC) problem. Different from \cite{Fowler81}-\cite{Ali11}, LSAA is a two-phase algorithm, in which the first phase allocates the sensor nodes into groups and the second phase searches a Set Cover (SC) for the GDC problem. The global SC consists of all local SCs that are found by a new local search algorithm called-Neighbor First Weighted Greedy Algorithm (NFWGA), with each SC covering one group. Here, the approximation ratio and the time complexity of the LSAA are studied rigorously. Additionally, a unique LSAA is proposed and evaluated to address the relay node Double Cover problem.

This paper then studies the NC problem that is NP-hard \cite{Lin99}-\cite{Dandekar12} and proposes a Relay Location Selection Algorithm (RLSA)-based NC algorithm. In this approach, RLSA first selects an optimal deployed location (nearest to the sink node in Euclidean distance) for each relay node, by utilizing the SC returned by LSAA, in building the high-tier network connectivity based on the Minimum Spanning Tree (MST) heuristic \cite{Lloyd07}. As a result of the RLSA outstanding performance, the high-tier network becomes denser and thus fewer relay nodes are added to the high-tier network for connectivity.

In summary, based on results and comparisons with the published results, this work is defining a new and important direction in relay node placement for two-tiered WSNs. Some of the significant contributions of this new approach are listed below:

\begin{itemize}
  \item Proposing a novel solution for the GDC problem, called Local Search Approximation Algorithm, which consists of two steps of first allocating the sensor nodes into independent groups and then proposing NFWGA to search a local SC for each group. Detailed and rigorous analysis on the approximation ratio and the time complexity of the LSAA are also provided in this paper.
  \item Proposing RLSA that selects one optimal deployed location for each relay node returned by LSAA. As a result, the RLSA-based NC algorithm saves newly added relay nodes for maintaining the network connectivity.
  \item Extensive simulations are conducted to demonstrate the efficiency of the algorithms proposed in this paper.
\end{itemize}

This paper is organized as follows. Section II reviews related work. Section III details the problem formulation. Section IV presents the approximation algorithm for the GDC problem. Section V presents the RLSA-based NC algorithm. Simulation results are presented in Section VI to demonstrate the superior performance of this work. Section VII finally concludes this paper.

\section{Related Work}
\subsection{GDC Problem}
In two-tiered WSNs, the placement of relay nodes is a significant factor in energy consumption and networks overhead. In order
to address the NP-complete nature of the relay node placement problem \cite{Fowler81}, existing research works mainly focus on three approaches of the Shift Strategy, the Grid Strategy and the Set-Covering Strategy. In this section, let $n$, $l$ and $\epsilon$ denote the number of sensor nodes, the shift parameter defined by the Shift Strategy and any positive constant, respectively.

\textbf{The Shift Strategy}: D. S. Hochbaum \cite{Hochbaum85} \cite{Hochbaum97} studied the GDC problem and proposed a Polynomial Time
Approximation Scheme (PTAS) based on the Shift Strategy, which divides the deployed field into stripes with the same width and partitions
these stripes into different groups by shifting the stripes, then solves the GDC problem in each stripe. The approximation ratio and time complexity of their scheme are bounded by $(1+1/l)^2$ and O$(l^2\lceil \sqrt{2}l\rceil^2n^{2\lceil \sqrt{2}l\rceil^2+1})$, with $l\geq 1$. T. Feder and D. Greene \cite{Feder88}, independently, Gonzalez \cite{Gonzalez91} considered the
related problem and proposed a $(1+1/l)$-approximation algorithm, which solves the problem by dividing the deployed field into stripes, with $l\geq 1$, and the time complexity of this algorithm is O$(6l\lceil \sqrt{2}l\rceil n^{6\lceil \sqrt{2}l\rceil+1})$. J. Tang et al. \cite{Tang06} proposed two approximation algorithms based on the PTAS proposed in \cite{Hochbaum85} \cite{Hochbaum97} to solve the relay node single cover problem and relay node double cover problem based on the Shift Strategy, with the approximation ratio of $4$ and $9/4$, respectively. A. Srinivas et al.
\cite{Srinivas09} considered the problem of the maintenance of a mobile backbone using the minimum number of backbone nodes, and they also
proposed algorithms for the GDC problem based on the PTAS proposed in \cite{Hochbaum85} \cite{Hochbaum97}, where the relay node had no mobility. However, the time complexity of the Shift Strategy grows exponentially with the shift parameter and the strategy is rather time-consuming when the shift parameter is large.

\textbf{The Grid Strategy}: M. Franceschetti et al. \cite{Franceschetti01} \cite{Franceschetti04} proposed a grid strategy, which divides
the deployed field into square meshes and the relay nodes can only be placed at the vertices of square meshes. They first analyzed the
approximation factor of grid strategy under different disc radius. Then they combined the Grid Strategy with the Shift Strategy, which
leads to a family of algorithm with a performance ratio of $\sigma(1+1/l)^2$, with $\sigma\in\{3,4,5,6\}$ and $l\geq 1$. The poor approximation ratio of the Grid Strategy stems from the fact that all placement locations in one grid are fixed.

\textbf{The Set-Covering Strategy}: In the Set-Covering Strategy, each relay node is represented by the sensor nodes covered by it, correspondingly,
the Set-Covering Strategy searches a minimum set cover of the given sensor nodes. Br\"{o}nnimamm and Goodrich \cite{Bronninamm95} studied the
set covering problem of dual VC-dimension and proposed an algorithm with a constant approximation ratio, which is not specified yet. Then
they gave several applications of their method to computational geometry including the GDC problem. K. N. Xu et al. \cite{Xu05} first modeled
GDC problem with a minimum set covering problem in WSNs, and proposed a recursive algorithm. K. Ali et al. \cite{Ali11} considered the related
problem and proposed a weighted greedy algorithm, with an approximation ratio of $\ln (n)+1$. Obviously, the Set-Covering Strategy yields an approximation ratio which grows logarithmically with the scale of the network and suffers from the scalability issue.

Unlike Set-Covering and Grid strategies that solve the whole GDC problem at a time, LSAA first decomposes the whole GDC problem into several GDC subproblems with smaller size (which is also different from the Shift Strategy that simply divides the deployed field into stripes), and then combines the solutions to each GDC subproblem to form the solution to the whole GDC problem. By the novel decomposition-based relay deployment method, LSAA has a small approximation ratio and a low time complexity.

\subsection{Network Connectivity Problem}
Extensive works have been done \cite{Tang06} \cite{Lin99}-\cite{Dandekar12} for the network connectivity problem and can be classified
into two branches, i.e., single-tiered network connectivity and two-tiered network connectivity.

In the single-tiered network, sensor nodes act as both transmitter and relay node. G. Lin and G. Xue \cite{Lin99} modeled the relay node placement
problem in the single-tiered network with the Steiner Minimum Tree with Minimum number of Steiner Points and bounded edge length (SMT-MSP) problem. The SMT-MSP problem was proved NP-hard and then solved by a $5$-approximation algorithm. Chen et al. \cite{Chen00} demonstrated that the algorithm proposed in \cite{Lin99} was actually a 4-approximation algorithm, and proposed a 3-approximation algorithm with time complexity O$(n^4)$, based on the so-called 4-star. Cheng et al. \cite{Cheng07} also studied the same problem of \cite{Chen00} and presented an improved 3-approximation algorithm with time complexity O$(n^3)$, based on the so-called 3-star. Misra et al. \cite{Misra08} \cite{Misra10} studied the constrained relay node placement problem in the single-tiered network and proposed a polynomial time O(1)-approximation algorithm.

In the two-tiered network, sensor nodes transmit the sensed packets to neighbor relay nodes and relay nodes forward the received packets to the sink node. Besides the GDC approximation algorithms, J. Tang et al. \cite{Tang06} also proposed two algorithms with approximation ratio of 8 and 4.5 for the connected relay node single cover problem and two algorithms with approximation ratio of 6 and 4.5 for the 2-connected relay node double cover problem, respectively.
E. Lloyd and G. Xue \cite{Lloyd07} first proposed a 7-approximation algorithm for the relay node placement problem in single-tiered network, and then presented a (5+$\epsilon$)-approximation algorithm for the relay node placement problem in two-tiered network. Q. Wang et al. \cite{Wang07} divided the relay node placement problem into two phases which correspond to the two problems in this paper, but mainly designed three heuristic algorithms to the network connectivity problem. D. J. Yang et al. \cite{Yang12} considered the constrained relay node placement problem in two-tiered network and presented two algorithms whose approximation ratios are O(1) and O(ln($n$)) for the connected single cover problem and 2-connected double cover problem, respectively. In \cite{Sandor08}-\cite{Dandekar12}, the relay node placement problem with the consideration of fault tolerance was studied. The main objective of these works was not only the high-tier network connectivity, but also the fault tolerance of the network (i.e., building at least two paths for each pair of sensor nodes).

The works in GDC have been one-sided that each work such as \cite{Tang06} \cite{Lin99}-\cite{Dandekar12} solve both GDC and NC problem separately. The critical flaw in these works is that they ignore the important relation between these two problems. In this work here, the novelty is that these two problems are considered jointly which will result in facilitating the work for the NC problem.

\section{Problem Formulation}
Let $X=\{x_1, x_2,\ldots, x_n\}$  denote a position set of \emph{n} sensor nodes, where $x_i$ ($1\leq i\leq n$) denotes the location of the \emph{i}th sensor node. Let $Y=\{y_1, y_2, \ldots, y_m\}$ denote a position set of \emph{m} relay nodes, where $y_i$ ($1\leq i\leq m$) denotes the location of the \emph{i}th relay node. Let a positive real number $r$ denote the communication radius of sensor nodes, and \emph{R} the communication radius of relay nodes. This paper assumes $R\geq 2r$.

Let \emph{p} and \emph{q} be two points in the plane. $\|p-q\|$ represents the Euclidean distance between \emph{p} and \emph{q}. If $\|x_i-y_j\|\leq r$, $1\leq i\leq n$ and $1\leq j\leq m$, then the sensor node $x_i$ can communicate with the relay node $y_j$, i.e., the sensor node $x_i$ is covered by the relay node $y_j$. If $\|x_i-x_j\|\leq 2r$, $1\leq i,j\leq n$, sensor nodes $x_i$ and $x_j$ are neighbors.

The RNC problem can be further divided into the Relay Node Single Cover (RNSC) problem and the Relay Node Double Cover (RNDC) problem.

\textbf{Definition 3.1 [Relay Node Single Cover (RNSC) problem]}. Given a set of sensor nodes $X=\{x_1, x_2,\ldots, x_n\}$, the RNSC problem searches a
set of relay nodes $Y=\{y_1, y_2, \ldots, y_m\}$, such that each sensor node $x_i$ $(1\leq i \leq n)$ is covered by at least one relay node.

To achieve a better network robustness, the RNDC problem is also considered in this paper and defined as follows.

\textbf{Definition 3.2 [Relay Node Double Cover (RNDC) problem]}. Given a set of sensor nodes $X=\{x_1, x_2,\ldots, x_n\}$, the RNDC problem searches a
set of relay nodes $Y=\{y_1, y_2, \ldots, y_m\}$, such that each sensor node $x_i$ $(1\leq i \leq n)$ is covered by at least two relay nodes.

\begin{figure}
\begin{center}
\includegraphics[height=3in,angle=270]{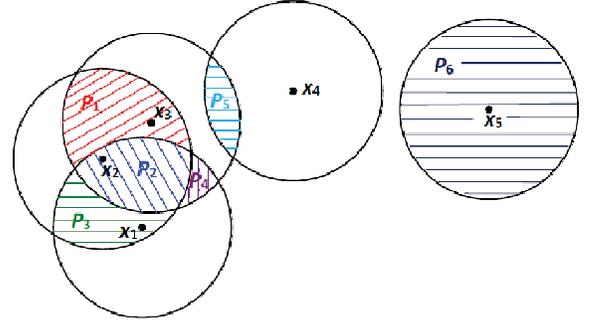}    
\caption{Possible positions for the set of sensor nodes $X=\{x_1, x_2, x_3, x_4, x_5\}$ are implied by the shades in different colors. The set of possible positions is $\mathcal{F} =\{P_1, P_2, P_3, P_4, P_5, P_6\}$.}
\label{fig1}                                 
\end{center}                                 
\end{figure}

The communication range of a sensor node $x_i$ can be described as a disk centered at location $x_i$ with the radius \emph{r}. In the RNSC problem, the potential locations of relay nodes should be selected in the first step. For a position set
 of sensor nodes $\tilde{X}$ ($1\leq |\tilde{X}|\leq n$), if there exists a point set $P$ such that $\|\tilde{x}-p\|\leq r$, $\forall p\in P$ and $\tilde{x}\in\tilde{X}$, then $P$ is considered as a \emph{possible position} to deploy relay nodes. For each sensor node having no neighbors, its communication range could be selected as a possible relay position. For ease of exposure, we describe each possible position as the set of the sensor nodes covered by it. For example, possible positions in Fig. \ref{fig1} can be formulated as $P_1=\{x_2, x_3\}$, $P_2=\{x_1, x_2, x_3\}$, $P_3=\{x_1, x_2\}$, $P_4=\{x_1, x_3\}$, $P_5=\{x_3, x_4\}$, $P_6=\{x_5\}$.

\begin{figure}
\begin{center}
\includegraphics[height=4.5cm]{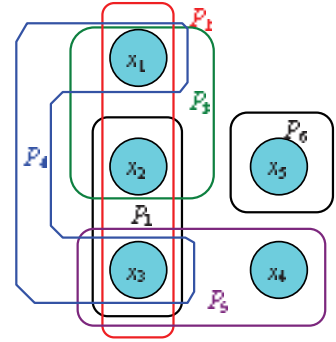}    
\caption{Formulating an example RNSC problem as a set covering problem.}
\label{fig2}                                 
\end{center}                                 
\end{figure}

In the second step, the RNSC problem searches a smallest subset which covers all sensor nodes, from the set of possible positions $\mathcal{F}$. In other words, the second step of RNSC problem reduces to the classic set covering problem for WSNs. As the example in Fig. \ref{fig2}, the RNSC problem with $X=\{x_1, x_2, x_3, x_4, x_5\}$ in Fig. \ref{fig1} is equivalent to searching for a smallest subset covering \emph{X} of $\mathcal{F}=\{P_1, P_2, P_3, P_4, P_5, P_6\}$.

\textbf{Definition 3.3 [Network connectivity problem]}. Given a set of relay nodes $Y_{GDC}$ distributed in the plane, place a smallest set of relay
nodes $Y_{NC}$ such that the undirected graph $G$ imposed on $Y=Y_{GDC}\cup Y_{NC}$ is connected.

\begin{figure}
\begin{center}
\includegraphics[height=4.5cm]{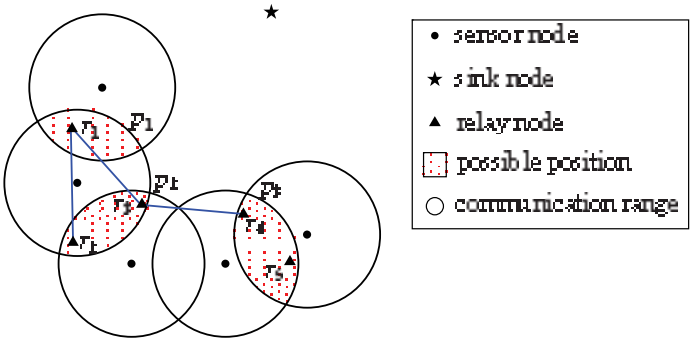}    
\caption{Impact of the location selection for each relay node.}
\label{fig3}                                 
\end{center}                                 
\end{figure}

The network connectivity problem is also known as the SMT-MSP problem
\cite{Lin99}-\cite{Dandekar12}. The RNSC problem returns a set of possible positions that form the backbone of the low-tier WSN. As a possible position is defined as the set of locations covering the same group of sensor nodes, we should select a specific location in each returned possible position to deploy one relay node. As shown in Fig. \ref{fig3}, $P=\{p_1, p_2, p_3\}$ is the set of returned possible positions and the blue line between two locations $r_i$ and $r_j$ indicates that $\|r_i-r_j\|\leq R$. For possible positions in $P$, if we choose locations $r_1$, $r_2$ and $r_5$ to deploy
relay nodes, then additional relay nodes (e.g., $r_3$ and $r_4$) will be needed to maintain the connectivity between $r_2$ and $r_5$. In contrast, if
we choose locations $r_1$, $r_3$ and $r_4$ to deploy relay nodes, then the connectivity of $r_1$, $r_3$ and $r_4$ straightforwardly holds. Therefore, the number of relay nodes deployed for the connectivity of high-tier WSN highly depends on the deployed location of relay nodes for covering the low-tier WSN. To save the relay nodes deployed for the connectivity of high-tier WSN, the deployed locations of possible positions returned by the RNSC problem should be optimized. Section V will focus on the optimization of deployed location in each possible position.

\section{Approximation Algorithms for RNSC and RNDC Problems}
\subsection{The Local Search Approximation Algorithm for RNSC Problem}
\textbf{Definition 4.1 [Neighbor of a possible position]}. For any two different elements $P_i$ and $P_j$ in $\mathcal{F}$, if at least one sensor node within $P_i$ (or $P_j$) is covered by $P_j$ (or $P_i$), $P_i$ and $P_j$ are mutual \emph{neighbors}.

\begin{figure}
\begin{center}
\includegraphics[height=5.5cm]{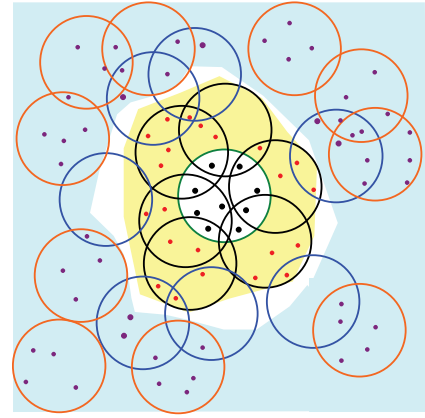}    
\caption{The LSAA illustration. The sensor nodes are represented by colorful points, and the coverage of relay nodes with different possible
positions are shown by circles with different colors.}
\label{fig4}                                 
\end{center}                                 
\end{figure}

Let $N(P_i)$ denote the set of neighbors of $P_i$ and $R_i=(P_1\cup P2\cup \ldots \cup P_k)-P_i$ denote the sensor nodes covered by $N(P_i)$ but not covered by $P_i$. As shown in Fig. \ref{fig4}, if $P_i$ corresponds to area covered by the green circle, then $N(P_i)$ represents the area covered by black circles and $R_i$ represents the set of the red points. Then, $X_i=X-P_i-R_i$ corresponds to the purple points located in blue region in Fig. \ref{fig4}.

\textbf{Definition 4.2 [Group]}. For any possible position, say $P_{i}$, the set of sensor nodes covered by the union of $P_{i}$ and $N(P_i)$ is defined as the \emph{group} of $P_{i}$, denoted by group $i$.

\textbf{Definition 4.3 [Bridge]}. The possible position covering sensor nodes of group $i$ and group $j$ simultaneously is called a \emph{bridge} between group $i$ and group $j$.

We denote $A^*$ as a Minimum Set Cover (MSC) of the set system $(X, \mathcal{F})$. The set consisting of all $bridge$s belonging to $A^*$ is denoted by $bridge_{A^*}$.

Given a set of sensor nodes $X=\{x_1,x_2,...,x_n\}$, let group $i$ (consisting of the red points and the black points in Fig. \ref{fig4}) denote an arbitrary group of $X$. As shown in Fig. \ref{fig4}, the bridges (the blue circles) of group $i$ cover sensor nodes both in group $i$ and other groups, i.e., group $i$ is associated with other groups by the bridges of group $i$. $A^*$ may include some of the bridges, if we know these bridges belonging to $A^*$ of group $i$ and remove these bridges, we just need to consider the possible positions (the black circles and the green circle in Fig. \ref{fig4}) which only cover group $i$ to cover group $i$. In other words, group $i$ becomes independent from other groups. Therefore, as we remove all the bridges (i.e., $bridge_{A^*}$) belonging to $A^*$ of each group, all the sensor nodes are divided into several independent groups. But this method is based on the assumption that the $bridge_{A^*}$ is known, and $bridge_{A^*}$ should be built based on $A^*$, whose solution is NP-complete. Hence, in the first phase of LSAA, we just group the sensor nodes without removing $bridge_{A^*}$.

The detail of the first phase of the LSAA is described as follows: LSAA starts with the multi-stage grouping phase. At each stage of the grouping phase, LSAA picks a possible position $P_i$
which covers the greatest number of remaining sensor nodes. Then, LSAA allocates the sensor nodes covered by $P_i$ and $N(P_i)$ as one group. The
grouping procedure above repeats until all sensor nodes are grouped.

Since fewer sensor nodes may lead to less searching time and favorable performance, we further classify the sensor nodes of group $i$ into two kinds, i.e.,
$R_i$ (the red points in Fig. \ref{fig4}) and $P_i$ (the black points in Fig. \ref{fig4}). Then, we search a SC $M_i$ for $R_i$. Let $w_i$ denote the number of sensor nodes, which are not covered by $M_i$, in $P_i$.
If $M_i$ fully covers $P_i$, we have $w_i=0$, otherwise, we have $w_i>0$. The second phase of LSAA is to carry out
a local search for each group. Specifically, LSAA checks the weight $w_i$ of $P_i$. If $w_i>0$, LSAA takes $\{P_i\}\cup M_i$ as the local search
result; if $w_i=0$, LSAA takes $M_i$ as the local search result. The whole procedure is repeated until all sensor nodes are covered. The LSAA is summarized in Algorithm \ref{LSAA}.

\begin{algorithm}[htb]         
\caption{Local Search Approximation Algorithm (LSAA).}             

\label{LSAA}                  

\begin{algorithmic}[1]                

\REQUIRE ~~\\                          

   A set of sensor nodes $X=\{x_1, x_2, \ldots, x_n\}$, wireless communication radius of sensor node \emph{r} and relay node \emph{R} $(R=2r)$ , and a set of possible positions $\mathcal{F}$.

\ENSURE ~~\\                           

    A set of relay nodes $Y=\{y_1, y_2, \ldots, y_m\}$, which ensures that each sensor node is covered by at least one relay node.

\STATE $PP=\varnothing$; \%$\varnothing$ is the empty set.
\STATE $U=X$;
\STATE $AP=\mathcal{F}$;
\WHILE{$(U\neq \varnothing)$}
\STATE Sort $AP$ in descending order according to the number of remaining sensor nodes covered by each possible position;
\STATE $P_i=$Pop($AP$); \% ``Pop" denotes a stack popping operation
\STATE find $N(P_i)$;
\STATE search $M_i$ by a local search approximation algorithm;
\STATE calculate $w_i$;
\STATE \textbf{switch}$(w_i)$
\STATE \quad  \textbf{case} $w_i>0$: $PP=\{P_i\}\cup M_i$;
\STATE \quad  \textbf{case} $w_i=0$: $PP= M_i$;
\STATE $j=1$;
\WHILE{($j<|PP|)$}
\STATE   $AP=AP-PP$;
\STATE   $TmpSens=$sensor nodes covered only by $PP_j$;
\STATE $k=1$;
\WHILE{($k<|AP|$)}
\IF{(($AP_k$ fully covers $TmpSens$) \&\& ($AP_k$ covers more sensor nodes than $PP_j$))}
\STATE $PP_j=AP_k$;
\ENDIF
\STATE $k=k+1$;
\ENDWHILE
\STATE $j=j+1$;
\ENDWHILE
\STATE $Tmp=$the sensor nodes covered by $PP$;
\STATE $U=U-Tmp$;
\STATE $AP=AP-PP$;
\STATE $Y=Y\cup PP$;
\ENDWHILE

\RETURN $Y$;                

\end{algorithmic}

\end{algorithm}

Before performing the performance analysis of LSAA, we have to first give out the definitions of some important symbols that will be used in the following proof.
 Let $\mathcal{C}_i$ and $A_i$ denote an MSC of group \emph{i} and the subset (covering the sensor nodes of group \emph{i}) of $A^*$, respectively. Let $B_i=A_i-bridge_{A^*}$ denote the set difference between $A_i$ and $bridge_{A^*}$. Obviously, $|A_i|\geq|\mathcal{C}_i|$ holds. Let $|OPA_i|$ denote the cardinality of a feasible SC (returned by the approximation algorithm) of group \emph{i}. Let $r_i=|OPA_i|/|\mathcal{C}_i|$ denote an approximation ratio of the local search algorithm at group \emph{i}. Let $r_{max}$ and $r_{min}$ denote the largest approximation ratio and the smallest approximation ratio, respectively, i.e., $r_{max}= \arg{\max\limits_{1\leq i\leq m}r_i}$ and ${r_{\min }} = \mathop {\arg \min }\limits_{1 \le i \le m} {r_i}$.
Let $D_i$ denote the set of possible positions (which cover the sensor nodes in group \emph{i}) of $bridge_{A^*}$.
Let $D_{max}$ denote the set with largest cardinality among $D_i$s, i.e., $|D_{max}|\in \arg{\max\limits_{1\leq i\leq m}|D_i|}$.

\begin{lemma}\label{lemma1}
 LSAA yields an approximation ratio of $(1+\bar{\epsilon})r_{max}$, where $\bar{\epsilon} =m|D_{max}|/|A^*|$.
\end{lemma}

\begin{IEEEproof}
According to the definition of $B_i$, we have
\begin{equation}\label{equ1}
    A_i=B_i\cup D_i, 1\leq i\leq m.
\end{equation}
Then, the following inequality (\ref{equ2}) holds
\begin{equation}\label{equ2}
    |B_i\cup D_i|\geq |\mathcal{C}_i|, 1\leq i\leq m,
\end{equation}
where (\ref{equ2}) straightforwardly follow $|A_i|\geq|\mathcal{C}_i|$.

As $\bigcup\limits_{i=1}^m|D_i|=bridge_{A^*}$, we have
\begin{equation}\label{equ3}
    \sum\limits_{i=1}^m|D_i|\geq \left|\bigcup\limits_{i=1}^mD_i\right|\geq|bridge_{A^*}|.
\end{equation}
Thus, there exists a non-negative constant \emph{C} such that
\begin{equation}\label{equ4}
    \sum\limits_{i=1}^m|D_i|=C+|bridge_{A^*}|.
\end{equation}
Then, we have
\begin{equation}\label{equ5}
 \begin{split}
  \sum\limits_{i=1}^m|B_i|+\sum\limits_{i=1}^m|D_i|&= \sum\limits_{i=1}^m|B_i|+|bridge_{A^*}|+C\\
        & \geq\sum\limits_{i=1}^m\left|B_i\bigcup D_i\right|\\
        &\geq_\alpha\sum\limits_{i=1}^m|\mathcal{C}_i|,
\end{split}
\end{equation}
where $"="$ and $"\geq_\alpha"$ follows from (\ref{equ4}) and (\ref{equ2}), respectively.

According to the definition of $A_i$, we can easily get
\begin{equation}\label{equ6}
    \sum\limits_{i=1}^m|B_i|+|bridge_{A^*}|=|A^*|.
\end{equation}

Plugging (\ref{equ6}) into (\ref{equ5}) yields
\begin{equation}\label{equ7}
    \begin{split}
      |A^*| &\geq\sum\limits_{i=1}^m|\mathcal{C}_i|-C  \\
        & \geq\sum\limits_{i=1}^m|\mathcal{C}_i|-\sum\limits_{i=1}^m|D_i|\\
        & \geq\sum\limits_{i=1}^m|\mathcal{C}_i|-m|D_{max}|.
    \end{split}
\end{equation}
or
\begin{equation}\label{euq8}
    \frac{m|D_{max}|}{|A^*|}+1\geq\frac{\sum\limits_{i=1}^m|\mathcal{C}_i|}{|A^*|}.
\end{equation}
Equivalently,
\begin{equation}\label{equ9}
    \frac{m|D_{max}|}{|A^*|}+1\geq\sum\limits_{i=1}^m\frac{|OPA_i|}{r_i|A^*|}\geq\frac{\sum\limits_{i=1}^m|OPA_i|}{r_{max}|A^*|}.
\end{equation}
Then, the approximation ratio of LSAA is given by
\begin{equation}\label{equ10}
    \frac{\sum\limits_{i=1}^m|OPA_i|}{|A^*|}\leq(\frac{m|D_{max}|}{|A^*|}+1)r_{max}=(1+\bar{\epsilon})r_{max},
\end{equation}
which completes the proof.
\end{IEEEproof}

\begin{remark} Inequality (\ref{equ10}) shows that the approximation ratio of LSAA only depends on $r_{max}$ and $\bar{\epsilon}$. Thus the approximation algorithm yielding small $r_{max}$ and $\bar{\epsilon}$ is favorable.
\end{remark}
\begin{lemma}\label{lemma2}
The ratio of $\frac{\sum\limits_{i=1}^m\left|OPA_i\right|}{|A^*|}$ has a lower bound of $r_{min}$.
\end{lemma}

\begin{IEEEproof}
According to the definition of $A^*$, we have

\begin{equation}\label{equ11}
\left|A^*\right|\leq \left|\bigcup\limits_{i=1}^m\mathcal{C}_i\right|\leq \sum\limits_{i=1}^m\left|\mathcal{C}_i\right|.
\end{equation}

According to the definition of $\mathcal{C}_i$, we have

\begin{equation}\label{equ12}
\left| {A^{*}} \right| \le \sum\limits_{i = 1}^m {\frac{{\left| {OP{A_i}} \right|}}{{{r_i}}}}  \le \sum\limits_{i = 1}^m {\frac{{\left| {OP{A_i}} \right|}}{{{r_{\min }}}}}.
\end{equation}

Then, we have

\begin{equation}\label{equ13}
\sum\limits_{i = 1}^m {\frac{{\left| {OP{A_i}} \right|}}{{\left| {A^{*}} \right|}}}  \ge {r_{\min }}.
\end{equation}
This completes the proof of Lemma \ref{lemma2}.
\end{IEEEproof}

\begin{remark}
Inequality (\ref{equ13}) implies that the only way to improve the lower bound of the performance of LSAA is to adopt local search algorithms with a low approximation ratio.
\end{remark}
\subsection{The Algorithm for Local Search}

\begin{figure}
\begin{center}
\includegraphics[height=3cm]{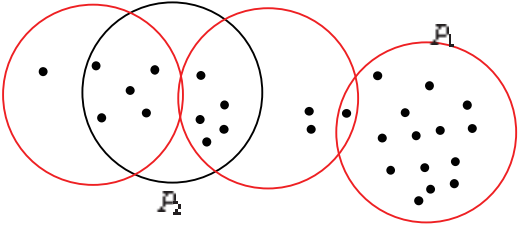}    
\caption{A case demonstrating the inefficiency of the Weighted Greedy algorithm. All the circles are returned by the Weighted Greedy algorithm, where the optimal solution only includes the red circles.}
\label{fig5}                                 
\end{center}                                 
\end{figure}

K. Ali et. al. \cite{Ali11} proposed a Weighted Greedy algorithm, which selects the possible position with the largest weight. The weight of possible position $P_i$ is defined as: $weight(P_i)=|U\cap P_i|-\alpha(|P_i|-|U\cap P_i|)$, $0<\alpha\leq 1/n$, where \emph{U} denotes the set of remaining sensor nodes. A case shown in Fig. \ref{fig5} demonstrates the inefficiency of the Weighted Greedy algorithm. Since the possible position $P_2$ has the largest weight after $P_1$ is selected, $P_2$ will be selected next and another two possible positions will be selected to realize the full cover to the sensor nodes. As a result, the SC returned by the Weighted Greedy algorithm includes 4 possible positions (all circles in Fig. \ref{fig5}), whereas the MSC for this case includes only 3 possible positions (red circles in Fig. \ref{fig5}). To avoid the inefficiency of Weighted Greedy algorithm, a Neighbor First Weighted Greedy Algorithm (NFWGA) is proposed.
The NFWGA does not only distinguish the remaining sensor nodes $U$ from the already covered sensor nodes, which are denoted by $X-U$, but also classify $U$ into two kinds of sensor nodes. The first kind of $U$ is denoted by $X_N$, which is covered by the neighbors of the deployed relay nodes. And the second kind of $U$ is denoted by $U-X_N$. The weights of sensor nodes of $X_N$ and $U-X_N$ are denoted by $\alpha$ and $\beta$, respectively. In order to increase the node degree of sensor nodes, we consider giving the sensor nodes in $X-U$ a weight $\gamma$, where the node degree of one sensor node is defined as the number of relay nodes covering this sensor node in this paper. The weight $\gamma$ is set far less than $\alpha$ and $\beta$ so as not to impact the size of the set cover of the remaining sensor nodes. The weight of possible position $P_i$ is calculated as
\begin{equation}\label{equ14}
    weight(P_i)=\delta\times \alpha+\varepsilon\times\beta+\zeta\times\gamma,
\end{equation}
where $\delta$, $\varepsilon$ and $\zeta$ denote the number of sensor nodes that are covered by $X_N$, $U-X_N$ and $X-U$, respectively. The proposed NFWGA is summarized in Algorithm \ref{NFWGA}.

\begin{algorithm}[htb]         
\caption{Neighbor First Weighted Greedy Algorithm (NFWGA).}             

\label{NFWGA}                  

\begin{algorithmic}[1]                

\STATE $P=\varnothing$; \%$\varnothing$ is the empty set.
\STATE $U=X$;
\STATE $Y=\varnothing$;
\STATE $AP=\mathcal{F}$;
\WHILE{$(U\neq \varnothing)$}
\STATE \emph{SN}= the remaining sensor nodes covered by the neighbor possible positions of \emph{Y};
\FORALL {$P_i\in AP$}
\STATE $weight(P_i)=\delta(SN)\times\alpha+\varepsilon(SN)\times\beta+\zeta(SN)\times\gamma$;
\ENDFOR
\STATE $P=\arg{\max\limits_{i=1,\ldots,m}weight(P_i)}$;
\STATE $U=U-P$;
\STATE $Y=Y\cup \{P\}$;
\STATE $AP=AP-\{P\}$;
\ENDWHILE

\RETURN $Y$;                

\end{algorithmic}

\end{algorithm}

\subsection{The Time Complexity Analysis of LSAA}

\begin{figure}
\begin{center}
\includegraphics[height=5cm]{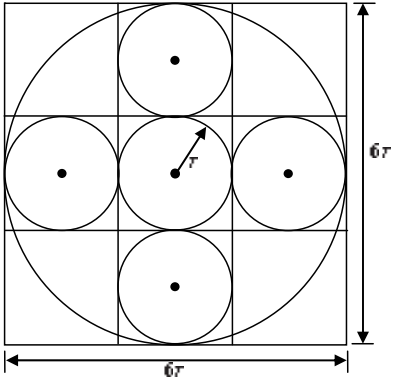}    
\caption{The illustration of the largest coverage of a group $i$. The largest coverage of $P_i$ and its neighbors is denoted by the big circle, which is an incircle of a square with side length of 6$r$. Obviously, the square can be divided into 9 small squares with side length of 2$r$. According to \cite{Hochbaum85} \cite{Tang06}, each square with side length of 2$r$ can be fully covered by 4 circles with radius $r$. This leads to that the square with side length 6$r$ can be fully covered by 36 circles with radius of $r$.}
\label{fig6}                                 
\end{center}                                 
\end{figure}

Let $t_i$ and $\bar{t}_i$ $(1\leq i \leq m)$ denote the running time of the NFWGA for group $i$ and the running time of grouping $i$, respectively. Then, the overall running time of LSAA ($T_{LSAA}$) is calculated as follows
\begin{equation}\label{equ15}
{T_{LSAA}} = \sum\limits_{i = 1}^m {\left( {{t_i} + {{\bar t}_i}} \right)} .
\end{equation}

The complexity of sorting AP as shown in step 5 of Algorithm \ref{LSAA} is O$(|\mathcal{F}|^2)$. Since the number of relay nodes to cover each group is no more than 36 as illustrated in Fig. \ref{fig6}, the complexity of the inner loop (steps 14-25) of Algorithm \ref{LSAA} is O$(|X|^2|\mathcal{F}|)$. As the time complexities searching for $P_i$ and $N(P_i)$ are O$(1)$ and O$(|X|^2|\mathcal{F}|)$, respectively, the time complexity of grouping $i$ is given by
\begin{equation}\label{equ16}
\begin{split}
{{\bar t}_i} &= \mathrm{O}({\left| {\cal F} \right|^2}) + \mathrm{O}({\left| X \right|^2}\left| {\cal F} \right|) + \mathrm{O}(1) + \mathrm{O}({\left| X \right|^2}\left| {\cal F} \right|)\\
 &= \mathrm{O}({\left| {\cal F} \right|^2}) + \mathrm{O}({\left| X \right|^2}\left| {\cal F} \right|).
 \end{split}
\end{equation}

For the given sensor nodes $X=\{x_1, x_2,...,x_n\}$, there are at most $|X|(|X|+1)$ possible positions, and thus we have $|\mathcal{F}|=\mathrm{O}(|X|^2)$, which together with (\ref{equ16}) implies that $\bar{t}_i=\mathrm{O}(|X|^4)$.

In NFWGA, the time complexity of calculating the weight of each possible position is $\mathrm{O}(|X|^2)$, and the time complexity of finding the possible position with maximal weight is $\mathrm{O}(|\mathcal{F}|)$. As the iteration number of the loop is less than $|X|$, the time complexity of NFWGA is given by
\begin{equation}\label{equ17}
{t_i} = \mathrm{O}({\left| X \right|^3}\left| {\cal F} \right|) = \mathrm{O}({\left| X \right|^5}).
\end{equation}

According to equation (\ref{equ15}), we have
\begin{equation}\label{euq18}
\begin{split}
T_{LSAA} &= \sum\limits_{i = 1}^m {\left( {\mathrm{O}({{\left| X \right|}^4}) + \mathrm{O}({{\left| X \right|}^5})} \right)} \\
 &= m\left( {\mathrm{O}({{\left| X \right|}^5})} \right).
\end{split}
\end{equation}

Since $m\leq |X|$, the time complexity of LSAA is finally given by $\mathrm{O}(|X|^6)$.

\subsection{The Algorithm for RNDC Problem}
To achieve robustness against the failure of deployed relay nodes, the RNDC problem is considered and the LSAA for Double Cover (LSAADC) is proposed to solve the RNDC problem. LSAADC consists of two steps. Specifically, in Step 1 the LSAA is run for the input sensor nodes, and in step 2 for the sensor nodes whose node degrees are less than 2 after step 1. The detail of LSAADC is shown in Algorithm \ref{LSAADC}.

\begin{algorithm}[htb]         
\caption{LSAA for Double Cover (LSAADC).}             

\label{LSAADC}                  

\begin{algorithmic}[1]                

\REQUIRE ~~\\                          
A set of sensor nodes $X=\{x_1, x_2, \ldots, x_n\}$, the transmission distance for sensor node and relay node is \emph{r} and relay node \emph{R} $(R=2r)$ , respectively, and a set $\mathcal{F}$ of all the possible position.

\ENSURE ~~\\                           
A set of relay nodes $Y=\{y_1, y_2, ... , y_m\}$, which makes each sensor node is covered by at least two relay node.

\STATE $RNSC=\varnothing$;
\STATE $Y=\varnothing$;
\STATE $SN=\varnothing$;
\STATE $RNDC=\varnothing$;
\STATE $RNSC=$the relay nodes returned by LSAA($X$);
\STATE $SN=$the sensor nodes covered by less than 2 relay nodes;
\STATE $RNDC=$the relay nodes returned by LSAA($SN$);
\STATE $Y=RNSC\cup RNDC$;
\RETURN $Y$;                

\end{algorithmic}

\end{algorithm}

\begin{remark} The worst case of LSAADC is that the node degree of each sensor node is 1 after the first step of LSAADC and all the sensor nodes are taken as the input of the second step. Therefore, the running time of LSASDC $T_{LSAADC}$ is upper bounded by $2T_{LSAA}$ and $T_{LSAADC}=\mathrm{O}({\left| X \right|}^6)$.
\end{remark}
\section{Network Connectivity Problem}
The proposed LSAA outputs a set of possible positions as the result for the given RNSC problem. However, a possible position includes numerous locations. Thus,
we should select an exact location for each returned possible position to deploy the relay node. Consequently, different selections will lead to distinct topologies of the high-tier WSNs, therefore, the quantity of relay nodes deployed to build the high-tier network connectivity varies with the selection. Thereby, the selection yielding a high-tier topology with fewer deployed relay nodes for the network connectivity is desired.

In \cite{Tang06} \cite{Ali11}-\cite{Dandekar12}, the intersection of two circles, whose centers are two sensor nodes and
radius is the communication radius of sensor nodes, is usually selected as the relay deployed location. Inspired by the fact that the denser network usually needs fewer relay nodes to build the network connectivity, this paper proposes a Relay Location Selection Algorithm (RLSA) to minimize the Euclidean distance between the deployed relay node and the sink node.  
As a result, the high-tier WSN shrinks towards the sink node and the average distance between any two nodes generally decreases (Although this conclusion does not hold mathematically, it has been verified by extensive simulations). Further, the transmission power and the number of newly added relay nodes for maintaining the connectivity of the high-tier WSN are reduced.


For the special case that one possible position only covers one sensor node, we can easily conclude that the nearest point in the disc (sensor's communication range) to the sink node is the intersection between the line segment (which connects the sink node and the sensor node) and the circle boundary of disc. However, when the special case does not hold, the relay location selection will be very complicated. In this section, we first consider a simple case that one possible position only covers two sensor nodes and then extend the result to the general case that one possible position covers more than two sensor nodes.
\subsection{A Simple Case}
\begin{figure}
\begin{center}
\includegraphics[height=5.5cm]{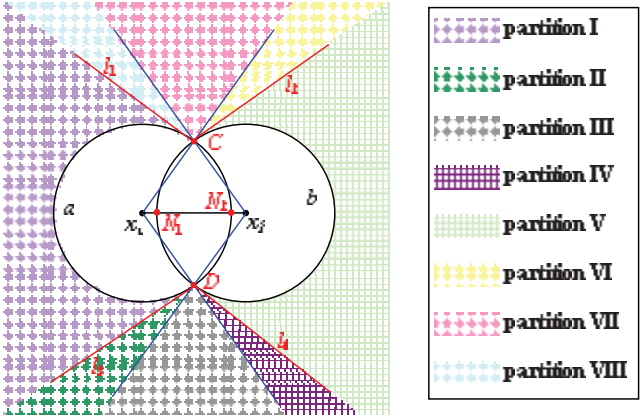}    
\caption{An illustration on the plane partitions out of the union of geometric discs $a$ and $b$.}
\label{fig7}                                 
\end{center}                                 
\end{figure}

Two sensor nodes are located at points $x_i$ and $x_j$ $(1\leq i,j\leq n)$, and $\|x_i-x_j\|\leq 2r$. The communication ranges of $x_i$
and $x_j$ are represented by discs $a$ and $b$, respectively. The edge circles of two discs intersect at points $C$ and $D$. Let $l_1$
and $l_2$ be the tangents of $a$ and $b$ at $C$, respectively. Let $l_3$ and $l_4$ be the tangents of $a$ and $b$ at $D$, respectively.
As shown in Fig. \ref{fig7}, the plane out of the union of $a$ and $b$ is divided into 8 partitions (i.e., partition $\mathrm{I-VIII}$) formed by lines
$l_1$, $l_2$, $l_3$, $l_4$, $x_jC$, $x_iC$, $x_jD$, and $x_iD$. The union of arcs $\wideparen{CN_1D}$ and $\wideparen{CN_2D}$ is termed as
the \emph{intersection arc} of $a$ and $b$, where $N_1$ and $N_2$ are the two intersections of line $x_ix_j$ and the \emph{intersection arc}.
Let $S$ denote the sink node which may be located in any partition from partition $\mathrm{I}$ to $\mathrm{VIII}$. Next, we will search for a point nearest to $S$ on the \emph{intersection arc}.
Let the angle between $l_1$ and $x_ix_j$ be denoted by $\alpha$, and $\angle Cx_jx_i$ be denoted by $\beta$ as shown in Fig. \ref{fig7}.

The following Lemma \ref{lemma3} and Lemma \ref{lemma4} tell that the nearest position to $S$ on the \emph{intersection arc} only depends on the location of $S$ and the size relationship between $\alpha$ and $\beta$. For this, four cases will be discussed in the proofs of Lemma \ref{lemma3} and Lemma \ref{lemma4}:

Case (1): the sink node $S$ lies in partition $\mathrm{I}$ or $\mathrm{V}$;

Case (2): the sink node $S$ lies in partition $\mathrm{II}$, $\mathrm{IV}$, $\mathrm{VI}$, or $\mathrm{VIII}$ ($\alpha < \beta$);

Case (3): the sink node $S$ lies in partition $\mathrm{VII}$ or $\mathrm{III}$;

Case (4): the sink node $S$ lies in partition $\mathrm{II}$, $\mathrm{IV}$, $\mathrm{VI}$, or $\mathrm{VIII}$ ($\alpha \geq\beta$).

\begin{lemma}\label{lemma3}
No matter Case (1) or Case (2) holds, 
the point nearest to $S$ on the \emph{intersection arc} is always the intersection of the line segment $x_jS$ or $x_iS$ and the \emph{intersection arc}.
\end{lemma}
\begin{IEEEproof}
We first consider Case (1) and assume that $S$ lies in the partition $\mathrm{I}$ (i.e., the area between tangents $l_1$ and $l_3$, as shown in Fig. \ref{fig8}).
For an arbitrary given point $T$ on $\wideparen{CN_2D}$ except $C$ and $D$, we know that the line segment $ST$ intersects $\wideparen{CN_1D}$ at one
point denoted as $T^*$. Obviously, $T^*$ is closer to $S$ than $T$. Therefore, the nearest point to $S$ must be on $\wideparen{CN_1D}$ and in the
rest of this proof we will only search for the nearest point to $S$ on $\wideparen{CN_1D}$.

\begin{figure}
\begin{center}
\includegraphics[height=5.5cm]{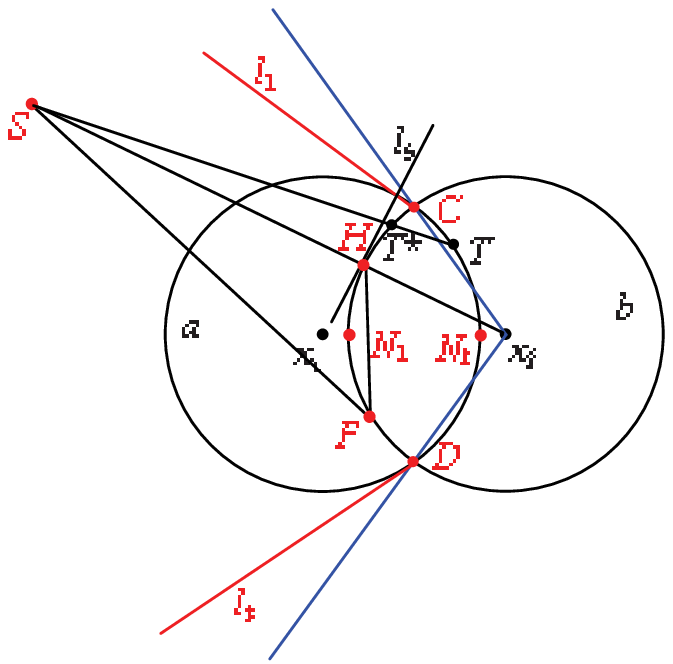}    
\caption{The illustration of the Case (1) in Lemma \ref{lemma3}.}
\label{fig8}                                 
\end{center}                                 
\end{figure}

Since the partition $\mathrm{I}$ is also located between the half lines $x_jC$ and $x_jD$, the line segment $x_jS$ intersects $\wideparen{CN_1D}$. Let $H$ be the intersection
of $x_jS$ and $\wideparen{CN_1D}$. Next we will show that $H$ is the point nearest to $S$ on $\wideparen{CN_1D}$.

Let $F$ be an arbitrary point on $\wideparen{CN_1D}$ except for $H$, and $l_5$ be the tangent of $b$ passing through $H$. Evidently, $l_5\perp x_jS$.
As $S$ and $F$ are located at the two sides of $l_5$, respectively, i.e., $\angle FHS>90^{\circ}$, according to the law of cosines, we know that $FS$ is
the longest side of $\triangle FHS$, i.e., $HS<FS$. In other words, $H$ is the nearest point to $S$ on the \emph{intersection arc}.
Because of the symmetry of discs, the case when $S$ lies in the partition $V$ can be similarly proved and thus skipped. So far, we have proved the first half of Lemma \ref{lemma3}.

\begin{figure}
\begin{center}
\includegraphics[height=4.5cm]{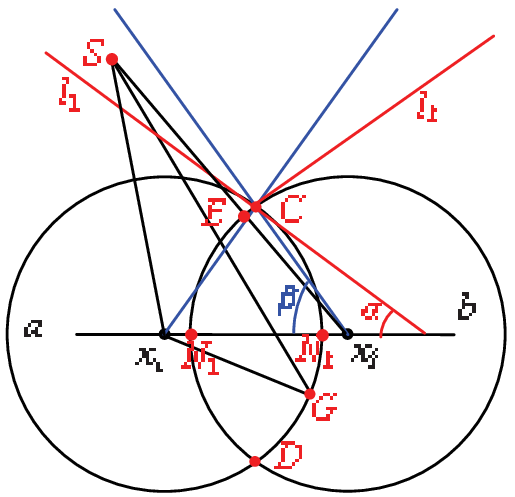}    
\caption{The illustration of the Case (2) in Lemma \ref{lemma3}.}
\label{fig9}                                 
\end{center}                                 
\end{figure}

Then we consider Case (2) and assume that $S$ lies in the partition $\mathrm{VIII}$ (i.e., the area between the half line $x_jC$ and $l_1$, as shown in Fig. \ref{fig9}). Let $G$ denote an arbitrary point on $\wideparen{CN_2D}$. Since $\alpha <\beta$ and $S$ lies in the partition VIII, we can see that $S$ lies above
the line $l_1$, which implies that $SG$ may intersect with $\wideparen{CN_2D}$ directly without intersecting with $\wideparen{CN_1D}$. Therefore, we cannot
prove the case (2) by straightly following the proof method of case (1).

Applying the law of cosines to $\angle Sx_iG$ and $\angle Sx_iC$, we have that
\begin{equation}\label{equ1-1}
\begin{array}{l}
S{C^2} = {r^2} + {x_i}{S^2} - 2r{x_i}S\cos \angle S{x_i}C\\
S{G^2} = {r^2} + {x_i}{S^2} - 2r{x_i}S\cos \angle S{x_i}G.
\end{array}
\end{equation}
Since the $\cos (x)$ decreases in $x$ when $0\leq x\leq \pi$, we conclude that $SC <SG$, which implies that $C$ is the nearest
point to $S$ on $\wideparen{CN_2D}$.

Since $\alpha <\beta$ and $S$ is located under the line $x_jS$,
obviously, $x_jS$ and $\wideparen{CN_1D}$ intersect at a point $E$. By the similar method in Case (1), we can prove that $E$ is the
nearest point to $S$ on $\wideparen{CN_1D}$, which implies that $E$ is closer to $S$ than $C$ and thus is the nearest point to
$S$ on the \emph{intersection arc}. Because of the symmetry of partitions, the case when $S$ is located in the partition $\mathrm{II}$, $\mathrm{IV}$ or $\mathrm{VI}$ ($\alpha <\beta$) can be similarly proved and thus skipped. This completes the proof of Lemma \ref{lemma3}.
\end{IEEEproof}

\begin{lemma}\label{lemma4}
No matter Case (3) or Case (4) holds, the point nearest to $S$ on the \emph{intersection arc} is always the point $C$ or $D$.
\end{lemma}

\begin{IEEEproof}
\begin{figure}
\begin{center}
\includegraphics[height=4.5cm]{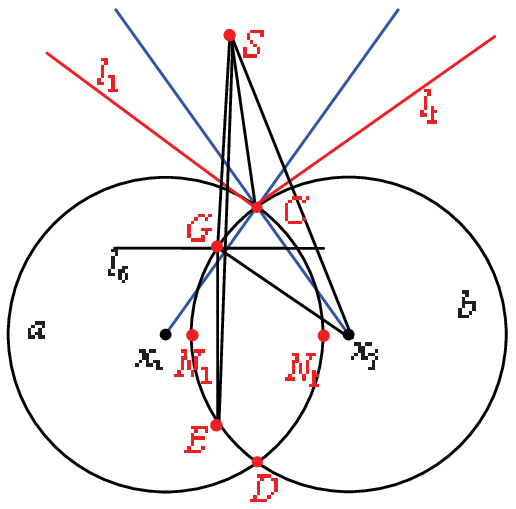}    
\caption{The illustration of the Case (3) in Lemma \ref{lemma4}.}
\label{fig10}                                 
\end{center}                                 
\end{figure}
First of all, we consider the Case (3). Again, because of the symmetry of discs, we only consider the case that $S$ is located in the
partition $\mathrm{VII}$ (i.e., the area between the half lines $x_jC$ and $x_iC$, as shown in Fig. \ref{fig10}).

Let $G$ denote an arbitrary point on the minor arc $\wideparen{CN_1}$ except for $C$. Let $l_6$ be a parallel line (passing through $G$)
of line $x_ix_j$, i.e., $l_6\parallel x_ix_j$. We then draw an orthogonal line $GE$ of $l_6$ at $G$ (i.e., $GE\perp l_6$), where $E$ is
the intersection of $GE$ and $\wideparen{DN_1}$. As $S$ and $E$ are located at the two sides of $l_6$, i.e., $\angle SGE > 90^{\circ}$,
according to the law of cosines, we know that $SE$ is the longest side of $\triangle SGE$, which implies that the nearest point to $S$ can
never on the minor arc $\wideparen{DN_1}$. Similarly, we can skip the minor arc $\wideparen{DN_2}$ and only search the nearest point to
$S$ on minor arcs $\wideparen{CN_1}$ and $\wideparen{CN_2}$.

Next we search the nearest point to $S$ from the minor arc $\wideparen{CN_1}$. Applying the law of cosines to $\angle x_jCS$ and $\angle x_jGS$
yields
\begin{equation}\label{equ19}
\begin{array}{l}
S{C^2} = {r^2} + {x_j}{S^2} - 2r{x_j}S\cos \angle S{x_j}C\\
S{G^2} = {r^2} + {x_j}{S^2} - 2r{x_j}S\cos \angle S{x_j}G.
\end{array}
\end{equation}
Since the $\cos(x)$ decreases in $x$ when $0\leq x\leq \pi$, we can conclude that $SC<SG$ since $\angle Sx_jC<\angle Sx_jG$, which implies
that $C$ is nearest point to $S$ on the minor arc $\wideparen{CN_1}$. Similarly, based on the law of cosines, we can achieve the conclusion that
$C$ is the nearest point to $S$ on the minor arc $\wideparen{CN_2}$. Thus, we have proved that $C$ is the nearest point to $S$ on the
\emph{intersection arc} when $S$ is located in the partition $\mathrm{VII}$. For the symmetry of discs, we can conclude that $D$ is the nearest point
to $S$ on the \emph{intersection arc} when $S$ lies at the partition $\mathrm{III}$.

\begin{figure}
\begin{center}
\includegraphics[height=4cm]{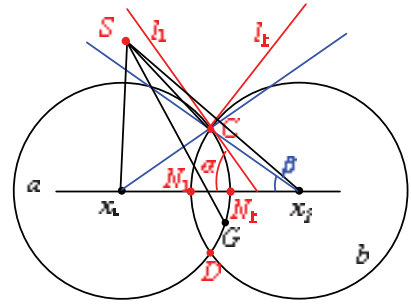}    
\caption{The illustration of the Case (4) in Lemma \ref{lemma4}.}
\label{fig11}                                 
\end{center}                                 
\end{figure}

Then, we consider the Case (4). We take the case when $S$ is located in the partition
$\mathrm{VIII}$ as shown in Fig. \ref{fig11} as an example. Since $\alpha \geq\beta$, partition $\mathrm{VIII}$ is the area between the half lines $x_jC$ and $x_iC$, which implies that neither $x_jS$
can intersect with $\wideparen{CN_1D}$ nor $x_iS$ can intersect with $\wideparen{CN_2D}$. Let $G$ be an arbitrary point except $C$ on the
\emph{intersection arc}. $SC<SG$ directly follows from the same proof for Case (3). Therefore, $C$ is the nearest point to $S$ on the
\emph{intersection arc} when $S$ lies in partition $\mathrm{VIII}$. Finally, for the symmetry of discs, the case when $S$ is located
in the partition $\mathrm{II}$, $\mathrm{IV}$, $\mathrm{VI}$ and $\alpha \geq \beta$, can be proved through the same method. This completes the proof of Lemma \ref{lemma4}.
\end{IEEEproof}

Combining Lemma \ref{lemma3} and Lemma \ref{lemma4}, we conclude that the nearest point to $S$ on the \emph{intersection arc} is the intersection point if either $x_iS$ intersects $\wideparen{CN_2D}$ or $x_jS$ intersects $\wideparen{CN_1D}$, and otherwise, the nearest point to $S$ on the \emph{intersection arc} is $C$ or $D$. The above analysis is summarized in Algorithm \ref{NPS}.

\begin{algorithm}[htb]         
\caption{Nearest Point to the Sink (NPS).}             

\label{NPS}                  

\begin{algorithmic}[1]                

\REQUIRE ~~\\                          

   A pair of discs $a$ and $b$ with radius $r$, centered at $x_1$ and $x_2$, respectively, $\|x_1-x_2\|\leq 2r$, the position of the sink node $S$.

\ENSURE ~~\\                           

    A point $p$ that is nearest to $S$ on the $intersection\ arc$ of $a$ and $b$.

\STATE $p=\varnothing$;
\STATE $cnt=1$;
\STATE find the intersections, which are denoted by $C$ and $D$, of $x_1$ and $x_2$; \%we denote the edge circles of $a$ and $b$ by $x_1$ and $x_2$,
respectively.
\WHILE{$(cnt<3)$}
    \STATE $p=$the intersection between $Sx_{cnt}$ and $x_{cnt}$;
    \IF{($p$ lies on the \emph{intersection arc})}
        \STATE \textbf{break};
    \ENDIF
    \STATE $p=\varnothing$;
    \STATE $cnt=cnt+1$;
\ENDWHILE
\IF{($p==\varnothing$)}
    \STATE $p=$ the point, which is nearest to $S$ among points $C$ and $D$;
\ENDIF
\RETURN $p$;                
\end{algorithmic}

\end{algorithm}

\subsection{The General Case}
This subsection extends the result of the simple case to the general case. The proposed RLSA is a multi-stage algorithm, with each stage consisting of three steps. In the first step of each stage, we randomly select one possible position $P$ and search the candidate locations (including the intersections and nearest point to $S$) on the intersection arc of each pair of sensors covered by $P$. Then, we select the locations fully covering $P$ among those candidate locations returned by the first step. Finally, we find a nearest location to $S$ by extensive search among these locations selected in the second step. The procedure above repeats until all returned possible positions are deployed. The RLSA is described in Algorithm \ref{RLSA}.


\begin{figure}
\begin{center}
\includegraphics[height=5.5cm]{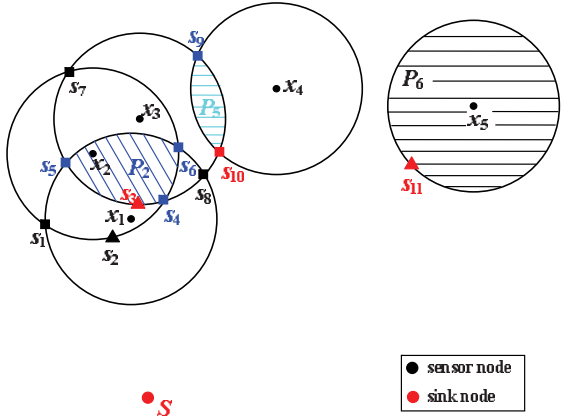}    
\caption{An example of the general case (four discs).}
\label{fig12}                                 
\end{center}                                 
\end{figure}

Fig. \ref{fig12} reloads the example in Fig. \ref{fig1} and uses it to illustrate how the RLSA works. It is evident form Fig. \ref{fig12} that $\{P_2, P_5, P_6\}$ is a possible position set (i.e., SC) that fully covers the sensor nodes $X=\{x_1, x_2, x_3, x_4, x_5\}$. First, we determine the deployed location of $P_2=\{x_1,x_2,x_3\}$. According to the RLSA, in the first step we search the intersections and nearest points for each pair of $x_1$, $x_2$, $x_3$, and these candidate locations are denoted by $s_1$, $s_2$, ..., $s_8$. In the second step, we find that only $\{s_3,s_4,s_5,s_6\}$ (denoted by the blue squares and red triangles) can fully cover $\{x_1, x_2, x_3\}$. Finally, we search the nearest point $s_3$ (denoted by the red triangle) to $S$ from $\{s_3,s_4,s_5,s_6\}$ by exhaustive search. As $P_5$ has only two candidate locations ($s_{10}$ and $s_{11}$), $s_{10}$ (denoted by the red square) is finally selected. Since $P_6$ has no neighbors, the intersection $s_{11}$ (denoted by the red triangle) of the line segment $x_5S$ and the circle centered at $x_5$ is selected as the deployed location of $P_6$. So far, we have selected locations $s_3$, $s_{10}$ and $s_{11}$ as the relay deployed locations of $P_2$, $P_5$ and $P_6$, respectively.


\begin{algorithm}[htb]         
\caption{Relay Location Selection Algorithm (RLSA).}             

\label{RLSA}                  

\begin{algorithmic}[1]                

\REQUIRE ~~\\                          

   A set of sensor nodes $X=\{x_1, x_2, \ldots, x_n\}$, a set of possible positions $P=\{p_1, p_2,...,p_m\}$, the location of the sink node $S$.

\ENSURE ~~\\                           

   A set of locations $L=\{l_1,l_2,...,l_m\}$ to deploy relay nodes.

\STATE $L=\varnothing$;
\WHILE{($P\neq\varnothing$)}
    \STATE $TmpP=$Pop($P$);
    \STATE $TmpX=$sensor nodes covered by $TmpP$ of $X$;
    \STATE $copyX=TmpX$;
    \WHILE{($TmpX\neq\varnothing$)}
        \STATE $Tmp1=$Pop($TmpX$);
        \STATE $neigh=$neighbor sensor nodes of $Tmp1$ in $TmpX$;
        \IF{($neigh=\varnothing$)}
            \STATE $PPTmp$=the nearest point to $S$ on the disc centered at $Tmp1$;
        \ENDIF
        \WHILE{($neigh\neq\varnothing$)}
            \STATE $Tmp2=$Pop($neigh$);
            \STATE $Tmp3=$NPS($Tmp1$, $Tmp2$, $S$);
            \STATE $Tmp4=$intersections of $Tmp1$ and $Tmp2$;
            \STATE $Tmp3=Tmp3\cup Tmp4$;
            \STATE $PPTmp=Tmp3\cup PPTmp$;
        \ENDWHILE
    \ENDWHILE
    \STATE $Tmp3=$points (of $PPTmp$) fully covering $copyX$;
    \STATE $Tmp4=$point (of $Tmp3$) nearest to $S$;
    \STATE $L=L\cup Tmp4$;
\ENDWHILE

\RETURN $L$;                

\end{algorithmic}

\end{algorithm}

In RLSA, the time complexity of the inner loop between step 12-18 is O($|X|$) and thus the complexity of the double loop between step 6-19 is O($|X|^2$). Since the iteration number of the main loop is $|P|$, the time complexity of RLSA is finally given by O($|P||X|^2$).

In this paper, we combine the RLSA with the MST heuristic proposed in \cite{Lloyd07} to build the network connectivity. Specifically, when the RNSC problem or RNDC problem is solved, we apply the RLSA to search an optimal deployed location for each possible position returned by the LSAA or LSAADC. Then the MST heuristic is employed to build the connectivity for the high-tier network by adding relay nodes if needed.
\section{Performance Evaluation}
In the simulation, sensor nodes are randomly deployed in a square area with the size $100\times 100$ m$^2$. The number of deployed sensor nodes $n$ ranges from 10 to 100. Communication radius for sensor node and relay node are $r=10$m and $R=20$m, respectively. The comparison results are averaged over 100 simulations. The simulation platform is developed by the C++, and simulations are executed on a 3.40-GHz Windows Work Station with 16 GB memory.

\subsection{Parameters Comparison for NFWGA}
\begin{figure}
\begin{center}
\includegraphics[height=5.8cm]{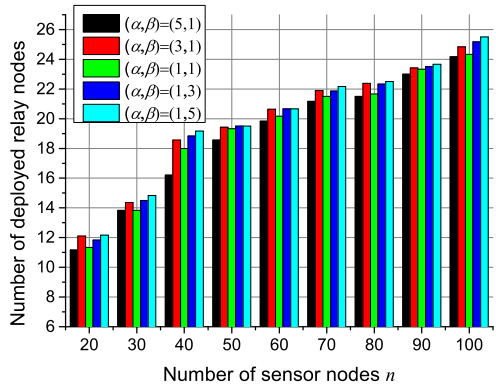}    
\caption{The influence of $\alpha$ and $\beta$.}
\label{fig13}                                 
\end{center}                                 
\end{figure}

Since the NFWGA is taken as the local search algorithm of LSAA, we first determine the value of parameters of NFWGA. In NFWGA the values of $\alpha$ and $\beta$ determine the size of set cover, and $\gamma$ is set quite small compared with $\alpha$ and $\beta$ so as not to influence the size of set cover. To be Specific, the exact value of $\gamma$ is set 0.01. For practical use, we obtain the value of $(\alpha, \beta)$ by simulation.
Fig. \ref{fig13} compares the performances of NFWGA under different values of $(\alpha, \beta)$. Comparison results show that the $(\alpha,\beta) =(5,1)$ yields the fewest deployed relay nodes. Without loss of generality, we set $(\alpha,\beta) =(5,1)$ in the following simulations.

\subsection{Performance Comparison for LSAA}
To demonstrate the efficiency of LSAA, the algorithms based on the Shift Strategy, the Grid Strategy and the Set-Covering Strategy are taken for comparison. For simplicity, in the rest of this paper, we denote the algorithm proposed in \cite{Tang06}, \cite{Franceschetti01} and \cite{Ali11} as Shift, Grid and Weighted
Greedy, respectively.

\begin{figure}
\begin{center}
\includegraphics[height=5.3cm]{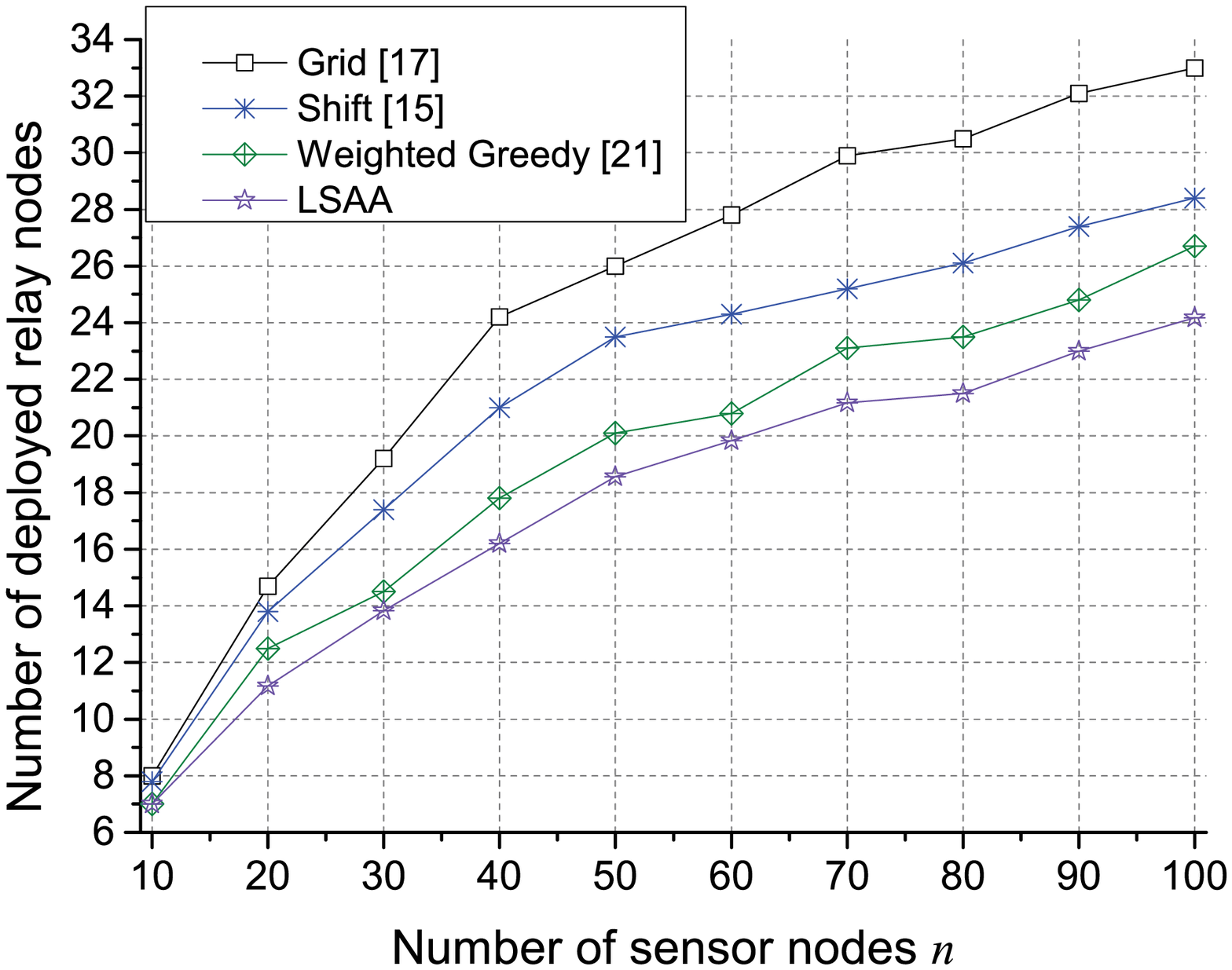}    
\caption{Comparisons on the number of deployed relay nodes} between LSAA and existing works.
\label{fig14}                                 
\end{center}                                 
\end{figure}

 Fig. \ref{fig14} compares the LSAA with \cite{Tang06}, \cite{Franceschetti01} and \cite{Ali11}. The local search phase of LSAA adopts the NFWGA.  It is clearly shown that LSAA returns the least number of relay nodes among all four algorithms. We observe that a large number of relay nodes could be saved due to LSAA in comparison to \cite{Tang06}, \cite{Franceschetti01} and \cite{Ali11}, in which the largest saving $(8/24\approx 33.3\%)$ occurs when $n=40$ and the Grid algorithm is adopted.

 \begin{figure}
\begin{center}
\includegraphics[height=5.8cm]{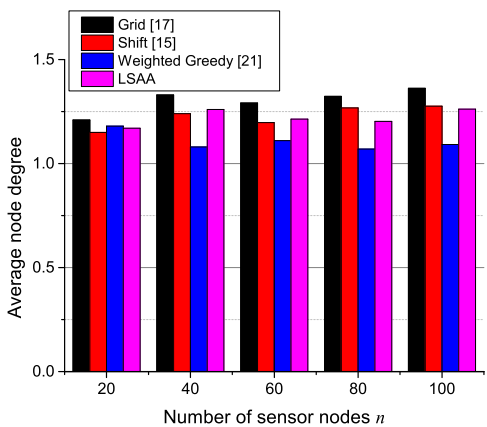}    
\caption{Comparison on the average node degree of sensor nodes between LSAA and existing works}.
\label{fig15}                                 
\end{center}                                 
\end{figure}

Fig. \ref{fig15} shows that the Grid algorithm is the most robust covering algorithm among the four algorithms. However, the poor performance in the number of deployed relay nodes limits the use of the Grid algorithm in practice. Although fewer relay nodes are deployed by LSAA, the average node degree of LSAA is generally larger than the Weighted Greedy algorithm, and comparable to the Shift algorithm as shown in Fig. \ref{fig15}, which can be explained by the introduction of weight $\gamma$ in NFWGA. To be specific, when multiple possible positions have the same weight $\delta\times \alpha+\varepsilon\times\beta$ on remaining sensor nodes, the possible position covering more covered sensor nodes will be selected.

 \begin{figure}
\begin{center}
\includegraphics[height=5.8cm]{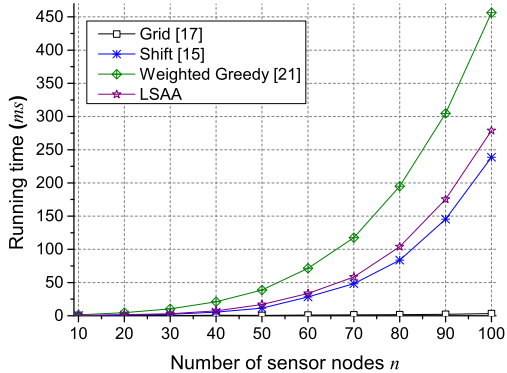}    
\caption{Comparison on the running time between LSAA and existing works.}
\label{figpp1}                                 
\end{center}                                 
\end{figure}

Fig. \ref{figpp1} shows that the running times of four compared algorithms increase as $n$ increases. LSAA has a moderate running time among the four algorithms. The largest running time of LSAA (happening when $n=100$) is around 275ms. Therefore, the time overhead of LSAA is affordable for practical use. 
\subsection{Performance Comparison for LSAADC}
\begin{figure}
\begin{center}
\includegraphics[height=5.8cm]{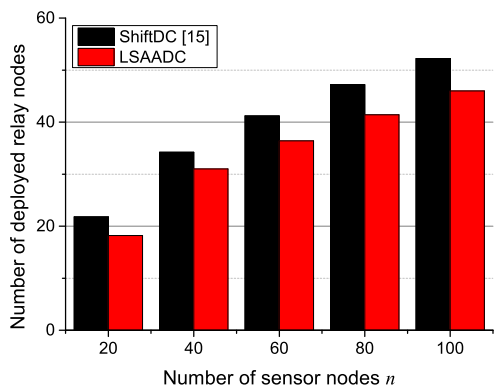}    
\caption{Comparison on the number of deployed relay nodes between LSAADC and ShiftDC.}
\label{fig16}                                 
\end{center}                                 
\end{figure}

\begin{figure}
\begin{center}
\includegraphics[height=5.8cm]{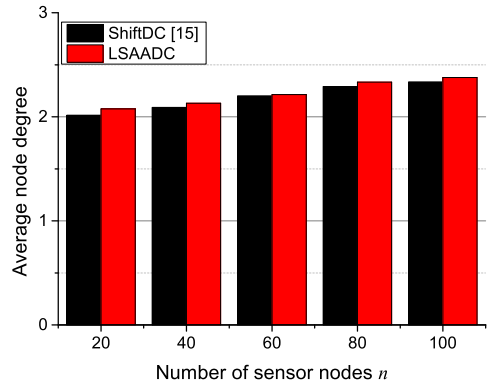}    
\caption{Comparison on the average node degree of sensor nodes between LSAADC and ShiftDC.}
\label{fig17}                                 
\end{center}                                 
\end{figure}

The algorithm proposed in \cite{Tang06} to the RNDC problem is denoted by ShiftDC in this paper. Fig. \ref{fig16} shows that LSAADC deploys fewer relay nodes for double cover than the ShiftDC. Meanwhile, in comparison to ShiftDC a slightly higher average node degree is achieved by LSAADC as shown in Fig. \ref{fig17}.

\begin{figure}
\begin{center}
\includegraphics[height=5.8cm]{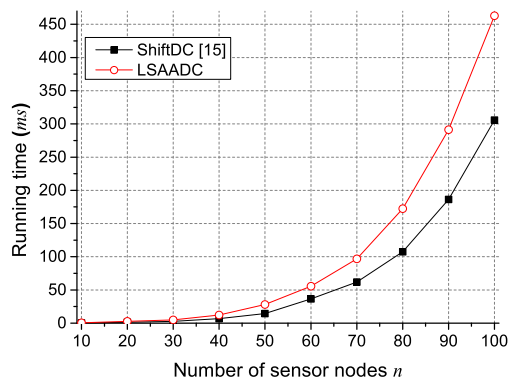}    
\caption{Comparison on the running time between LSAADC and ShiftDC.}
\label{figpp2}                                 
\end{center}                                 
\end{figure}

In addition, the comparison on running time between LSAADC and ShiftDC is shown in Fig. \ref{figpp2}. Similar to the comparison between LSAA and the Shift algorithm in Fig. \ref{figpp1}, ShiftDC also outperforms LSAADC and the performance gap between ShiftDC and LSAADC becomes more visible in comparison to Fig. \ref{figpp1}.

\subsection{Performance Comparison for RLSA}
\begin{figure}
\begin{center}
\includegraphics[height=5.8cm]{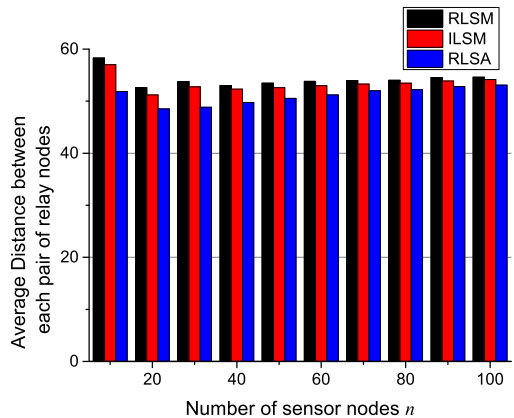}    
\caption{Comparison on the average distance between each pair of deployed relay nodes.}
\label{fig18}                                 
\end{center}                                 
\end{figure}

In order to examine the effectiveness of RLSA, we first show that the high-tier network will become dense when we perform RLSA to select relay deployed locations. Fig. \ref{fig18} compares RLSA with the Intersection Location Selection Method (ILSM) \cite{Tang06} \cite{Ali11} and a Random Location Selection Method (RLSM) in terms of the average distance between each pair of deployed relay nodes. We can see from Fig. \ref{fig18} that the RLSA returns the smallest average distance among all three algorithms and RLSM returns the largest average distance.

\begin{figure}
\begin{center}
\includegraphics[height=5.8cm]{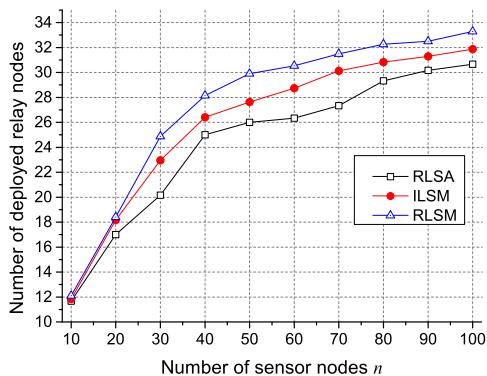}    
\caption{Comparison on the average number of deployed relay nodes between RLSA-based, ILSM-based and RLSM-based connectivity algorithms.}
\label{fig20}                                 
\end{center}                                 
\end{figure}

Fig. \ref{fig20} then presents the comparison on the average number of deployed relay nodes between RLSA-based, ILSM-based and RLSM-based connectivity algorithms. Notice that, for fair comparison, all of the three compared algorithms adopt an MST-based algorithm-Feasible single-tiered Relay Node Placement (F1tRNP) \cite{Lloyd07} to solve the network connectivity problem. As RLSA selects the nearest deployed position for each relay node, the average distance between relay nodes and the sink node of RLSA will be the smallest. In other words, because of RLSA the high-tier network will shrink towards the sink node. This explains that RLSA-based network connectivity algorithm needs fewer relay nodes for network connectivity than the other two algorithms. It is shown in Fig. \ref{fig20} that the RLSA-based connectivity algorithm at most saves $13.3\%$ deployed relay nodes comparing with other algorithms when $n=60$.

\begin{figure}
\begin{center}
\includegraphics[height=5.8cm]{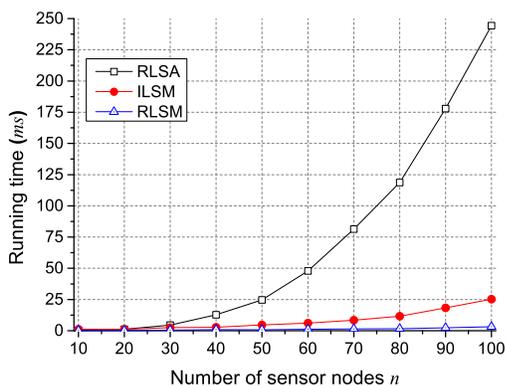}    
\caption{Comparison on the running time between RLSA, ILSM and RLSM.}
\label{figpp3}                                 
\end{center}                                 
\end{figure}

Fig. \ref{figpp3} shows the comparison on the running time of RLSA, ILSM and RLSM. It is clearly shown that as $n$ increases, the running time of RLSA grows much faster than those of ILSM and RLSM, which can be explained by the fact that the inner loop (Step 6-19) of Algorithm 5 (RLSA) contributes an additional time complexity $\mathrm{O}(n^{2})$.



\begin{figure}
\begin{center}
\includegraphics[height=5.8cm]{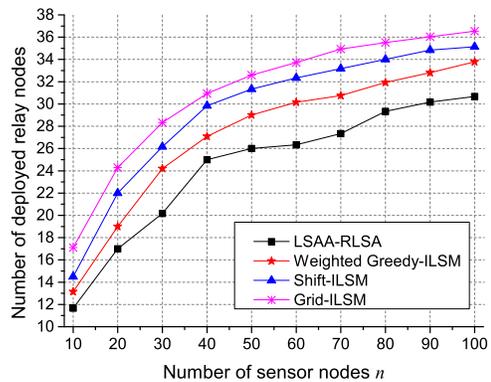}    
\caption{Comparison on the average number of deployed relay nodes between LSAA-RLSA, Shift-ILSM, Grid-ILSM} and Weighted Greedy-ILSM.
\label{fig21}                                 
\end{center}                                 
\end{figure}

\begin{figure}
\begin{center}
\includegraphics[height=5.8cm]{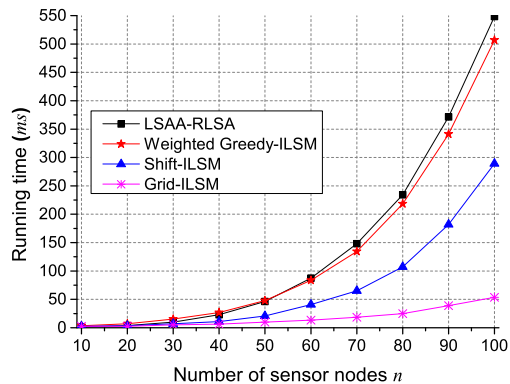}    
\caption{Comparison on the running time between LSAA-RLSA, Shift-ILSM, Grid-ILSM and Weighted Greedy-ILSM.}
\label{figpp4}                                 
\end{center}                                 
\end{figure}

Finally, in order to evaluate the overall performance of this work (denoted by LSAA-RLSA, which consists of LSAA and the RLSA-based network connectivity algorithm), the algorithm presented in \cite{Tang06}, \cite{Franceschetti01} and \cite{Ali11}, which are denoted by Shift-ILSM,
Grid-ILSM and Weighted Greedy-ILSM, respectively, are taken for comparison. All of the three compared algorithms employ F1tRNP \cite{Lloyd07} to solve the network connectivity problem. Fig. \ref{fig21} and Fig. \ref{figpp4} show the number of deployed relay nodes and running time of these algorithms, respectively. To sum up, LSAA-RLSA is the most time-consuming but at the same time also the most cost-effective algorithm. As the largest running time of LSAA-RLSA in simulation is 550ms (which is still affordable for practical use), the proposed LSAA-RLSA is the most favorable one among the four algorithms because of its superiority in the cost of relay deployment.

\section{Conclusion}
In this paper, we have investigated the relay node placement problem in WSNs. Firstly, by modeling the relay node placement problem as a GDC problem, we proposed a novel algorithm-LSAA to solve it. The novelty of LSAA lies in the separation of the grouping phase and the local search phase. The LSAA yields an approximation ratio of $(1+\bar{\epsilon})r_{\max}$, and the time complexity of LSAA is proved to be O$(|X|^6)$, where $r_{\max}$ denotes the maximal local approximation ratio. The number of deployed relay nodes of LSAA is compared with the existing works through simulations and the results show that at most one third of relay nodes could be saved due to this new approach of LSAA. Secondly, in order to improve the network robustness against connection failure, the RNDC problem was solved to enhance the node degree of the network. The efficiency of the solution algorithm-LSAADC to the RNDC problem is demonstrated through simulations. The simulation results show that the LSAADC deploys fewer relay nodes while achieving a higher node degree in comparison to existing works. Thirdly, a unique RLSA-based NC algorithm was proposed to search a nearest deployed location to the sink node for each relay node. Then the MST heuristic is employed to build the connectivity for the high-tier WSN. Again, the number of newly added relay nodes for building the connectivity of the high-tier WSN is significantly saved in simulations. In summary, based on comparisons with the published works and the presented results, this work outlines a new path for relay node placement for the new generations of WSN applications.



%

\appendices


%
%

\ifCLASSOPTIONcaptionsoff
  \newpage
\fi

\end{document}